\documentclass[twocolumn,aps,prl,10pt,superscriptaddress]{revtex4-1} 

\usepackage{graphicx}
\usepackage{bm}
\usepackage{dcolumn}
\usepackage{amssymb}
\usepackage{amsmath}
\usepackage{amsbsy}
\usepackage{amsfonts}
\usepackage{epsfig}
\usepackage{graphicx}
\usepackage{multirow}
\usepackage{stackrel}
\usepackage{paralist}
\usepackage[rgb]{xcolor}
\usepackage[textsize=tiny,textwidth = 1.6 cm]{todonotes}

\usepackage[export]{adjustbox}

\usepackage[percent]{overpic}
\usepackage{tikz}
\usetikzlibrary{intersections} 
\usetikzlibrary{positioning,calc}
\usepackage[caption=false]{subfig}

\usepackage[normalem]{ulem}

\tikzset{
    right angle quadrant/.code={
        \pgfmathsetmacro\quadranta{{1,1,-1,-1}[#1-1]}     
        \pgfmathsetmacro\quadrantb{{1,-1,-1,1}[#1-1]}},
    right angle quadrant=1, 
    right angle length/.code={\def\rightanglelength{#1}},   
    right angle length= 1 ex, 
    right angle symbol/.style n args={3}{
        insert path={
            let \p0 = ($(#1)!(#3)!(#2)$) in     
                let \p1 =  ($(\p0)!\quadranta*\rightanglelength!(#3)$),
                \p2 = ($(\p0)!\quadranta*\rightanglelength!(#2)$) in 
                let \p3 = ($(\p1)+(\p2)-(\p0)$) in  
            (\p1) -- (\p3) -- (\p2)
        }
    }
}

\newcommand{\cancel}[1]{}

\newcommand{\I}{\mathrm{i}}

\DeclareMathOperator{\Tr}{Tr}

\DeclareMathOperator{\sinc}{sinc}

\newcommand{\refEq}[1]{Eq.~(\ref{#1})}
\newcommand{\refFig}[1]{Fig.~\ref{#1}}
\newcommand{\refSctn}[1]{Section~\ref{#1}}

\newcommand{\citeRef}[1]{Ref.~[\onlinecite{#1}]}

\definecolor{NewColor}{rgb}{1,0,0}
\definecolor{myRed}{rgb}{1,0,0}
\definecolor{myGreen}{rgb}{0.2,0.6,0.2}
\definecolor{myBlue}{rgb}{0,0,1}

\graphicspath{}

\newcommand{\CO}[1]{\textcolor{red}{}}

\newlength{\figHeight}
\newlength{\noLengthl}

\renewcommand{\vec}[1]{{\boldsymbol{#1}}}

\usepackage{amsmath,amssymb,amsfonts,graphicx,multirow,color,bm}
 
\newcommand{\bee}{\begin{eqnarray}}
\newcommand{\ee}{\end{eqnarray}}
\newcommand{\bma}{\begin{pmatrix}}
\newcommand{\ema}{\end{pmatrix}}
\newcommand{\balig}{\begin{align}}
\newcommand{\ealig}{\end{align}}

\newcommand{\ba}{\begin{align}}
\newcommand{\ea}{\end{align}}

\newcommand{\ignore}[1]{}

\newcolumntype{C}[1]{>{\centering\let\newline\\\arraybackslash\hspace{0pt}}m{#1}}

\begin{document}
\setlength{\figHeight}{25ex}
\title{Vortex Majorana braiding in a finite time}
\author{Thore Posske}
\affiliation{I. Institut f{\"u}r Theoretische Physik, Universit{\"a}t Hamburg, Jungiusstra{\ss}e 9, 20355 Hamburg, Germany}
\author{Ching-Kai Chiu}
\affiliation{Kavli Institute for Theoretical Sciences, University of Chinese Academy of Sciences, Beijing 100190, China}
\author{Michael Thorwart}
\affiliation{I. Institut f{\"u}r Theoretische Physik, Universit{\"a}t Hamburg, Jungiusstra{\ss}e 9, 20355 Hamburg, Germany}
\begin{abstract}
Abrikosov vortices in Fe-based superconductors are a promising platform for hosting Majorana zero modes. Their adiabatic exchange is a key ingredient for Majorana-based quantum computing.
However, the adiabatic braiding process 
can not be realized in state-of-the-art experiments.
We propose to replace the infinitely slow, long-path braiding by only slightly moving vortices in a special geometry without actually physically exchanging the Majoranas, like a Majorana carousel. 
Although the resulting finite-time gate is not topologically protected, it is robust against variations in material specific parameters and in the braiding-speed. We prove this analytically. Our results carry over to Y-junctions of Majorana wires.
\end{abstract}

\maketitle

\paragraph{Introduction.}
Recent experiments on low-dimensional superconducting structures 
have revealed localized electronic states at the Fermi level. Although still debated, these states can be attributed to 'half-fermionic' exotic electronic states, the Majorana zero modes.
Spatially isolated Majorana zero modes are not lifted from zero energy when coupled to ordinary quasiparticles and are a key ingredient for demonstrating nonuniversal topological quantum computing \cite{Kitaev2001UnpairedMajoranaFermions,Alicea2011NonAbelianStatisticsIn1DWireNetworks,ElliotFranz2015,NayakSimonSternFreedmanDasSarma2008NonAbelianAnyonsAndTopologicalQuantumComputing}, despite some susceptibility to external noise \cite{BudichWalterTrauzettel2012FailureOfProtectionOfMajoranaBasedQubitsAgainstDecoherence,PedrocchiDiVincenzo2015MajoranaBraidingWithThermalNoise}.
Majorana modes have supposedly been detected at the ends of semiconducting wires \cite{MourikKouwernhoven2012SignaturesOfMajoranaFermionsInHybridSCSMNanowireDevices,DasRonenMostOregHeiblumShtrikman2012ZeroBiasPeaksAndSplittingInAnAlInAsNanowireTopologicalSuperconductor}, 
designed atomic chains with helical magnetic structures  \cite{KimPalacioMoralesPosskeEtAl2018TowardTailoringMajoranaBoundStatesInArtificiallyConstructedMagneticAtomChainsOnElementalSuperconductors}, and, in particular, at the surface of superconductors with a superficial Dirac cone, e.g., Fe-based superconductors. There, Abrikosov vortices carry spatially localized peaks in the density of states at zero bias voltage \cite{Ivanov2001NonAbelianStatisticsOfHalfQuantumVorticesInPWaveSuperconductors,FuKane2008SuperconductingProximityEffectAndMajoranaFermionsAtTheSurfaceOfATopologicalInsulator,Wang..Gao2018EvidenceForMajoranaBoundStatesInAnFeBasedSC,
Yin..Pan2015ObservationOfARobustZeroEnergyBoundStateInFeBasedSC,
Liu..Feng2018RobustAndCleanMajoranaZeroModeInTheVortexCoreOfHighTCSCLiFeOHFeSe,Zhang182,Machida:2019aa,Yuan..Xue2019EvidenceOfAnisotropicMajoranaBoundStatesIn2MWS2,ZhuEtAl2020NearlyQuantizedConductancePlateauMajorana,ChiuEtAl2020HybridizingMajoranaVorticesAtLargerField}.
We call the latter vortex Majoranas. 

The next milestone towards topological quantum computing as well as the final evidence for the existence of Majorana zero modes is to achieve Majorana braiding, i.e., moving two Majorana zero modes around each other 
adiabatically 
 \cite{NayakSimonSternFreedmanDasSarma2008NonAbelianAnyonsAndTopologicalQuantumComputing,KarzigOregRefaelFreedman2018RobustMajoranaMagicGatesViaMeasurements,KarzigKnappLutchynBondersonHastingsNayakAliceaFlensbergPluggeOregMarcusFreedman2017ScalableDesignsForQuasiparticlePoisoningProtectedTopologicalQuantumComputingWithMajoranaZeroModes,KarzigOregRefaelFreedman2016UniversalGeometricPathToARobustMajoranaMagicGate, Alicea2011NonAbelianStatisticsIn1DWireNetworks,Halperin2010AdiabaticManipulationsOfMajoranaFermionsInA3DNetworkOfQuantumWires,Alicea2011NonAbelianStatisticsIn1DWireNetworks, VijayHsiehFu2015MajoranaFermionSurfaceCodeForUniversalQuantumComputations,KarzigOregRefaelFreedman2016UniversalGeometricPathToARobustMajoranaMagicGate}.
This naive implementation of Majorana braiding in Fe-based superconductors poses major experimental problems despite the fundamental limitation that a true adiabatic evolution of perfectly degenerate levels cannot be achieved in principle. 
First, the length of the exchange path is on the order of micrometers such that braiding would take up to minutes in current setups, introducing high demands on sample quality, temperature, and experimental control for guaranteeing coherent transport of the vortex without intermediate quasiparticle poisoning \cite{NovemberSauWilliams2019SchemeForMajoranaManipulationUsingMFS}. Second, braiding the vortices results in twisted flux lines in the bulk of the Fe-based superconductor.
This induces an energetic instability  \cite{Roy..Raychaudhuri2019MeltingOfTheVortexLatticeThroughIntermediateHexaticFluidInAnAlphaMoGeThinFilm}, hindering braiding and eventually causing relaxation events that disturb the zero-energy subspace and thereby quantum computation.

\begin{figure*}
\centering
\includegraphics[width = \linewidth]{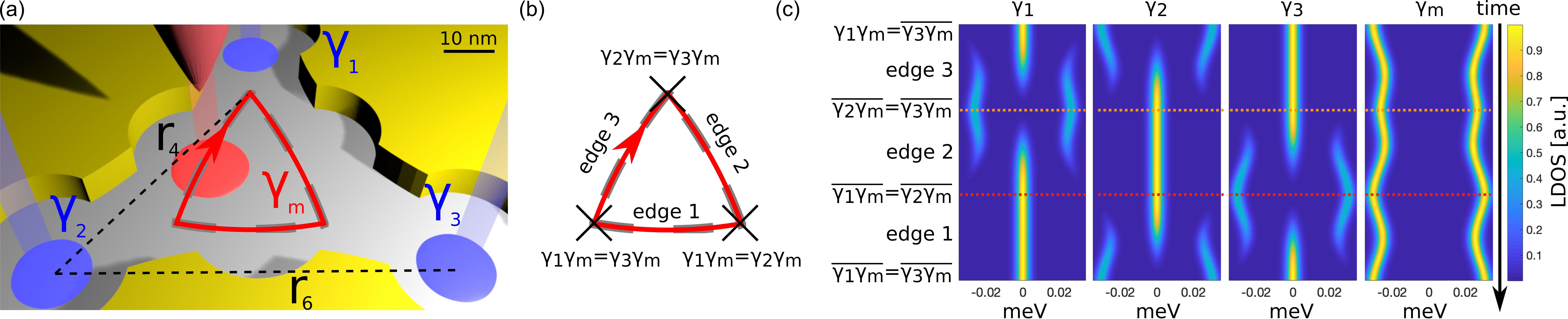}
\caption{\label{figSetup}\label{figSetup_path}\label{LDOS_evolution_new}
{(a)} Setup: Three decoupled vortex Majoranas $\gamma_1$, $\gamma_2$, $\gamma_3$ (blue) that are $r_6$ away from each other. A fourth, movable vortex Majorana $\gamma_{\text{m}}$ (red) is on a path that is $r_4$ away from the exterior ones.
{(b)}
Path of $\gamma_{\text{m}}$ to realize the braiding gate.
{(c)}
LDOS evolution for each vortex core as $\gamma_{\text{m}}$ is moved.
Two zero-bias peaks at exterior vortices and an energy splitting at the movable vortex are always present. The LDOS for each vortex does not break particle-hole symmetry significantly, supporting the validity of a low-energy model.
}
\end{figure*}

In this manuscript, we substantially simplify the direct approach of physically braiding vortex Majoranas and show how 
braiding is realized within a finite time.
To this end, vortex Majoranas are spatially arranged such that changing the position of one of them on a short, well-defined path is equivalent to ordinary braiding.
For current experimental systems like FeTe$_x$Se$_{1-x}$ \cite{Yin..Pan2015ObservationOfARobustZeroEnergyBoundStateInFeBasedSC,KunDaiWang2019QuantumAnomalousVortexAndMajoranaZeroModeInFeBasedSC,PhysRevB.81.180503,PhysRevB.82.184506}, the allowed time scales for our braiding operation ranges from adiabatically slow up to nanoseconds.
The protocol is robust against variations in material parameters and in the local speed of the vortex motion, which we prove by an analytical finite-time solution. 
Unwanted couplings that lift the degeneracy of the ground state are excluded by the special spatial arrangement of the vortices and additionally exponentially suppressed on the Majorana superconducting coherence length $\xi$.

For our proposal, we presume that a reliable mechanism  for moving vortices is available. Tremendous progress in this regard has recently been made by moving vortices with the cantilever of a magnetic force microscope \cite{doi:10.1063/1.3000963,NovemberSauWilliams2019SchemeForMajoranaManipulationUsingMFS,PolshynNaibertBudakian2019ManupulatingMultiVortexStatesInSuperconductingStructures}. 
Additionally, the controlled nanoscale assembly of vortices has been achieved with a heated tip of a scanning tunneling microscope (STM) \cite{Ge..Moshchalkov2016NanoscaleAssemblyOfSCVorticesWithScanningTunnelingMicroscopeTip} by letting the vortices follow the locally
heat-suppressed
superconducting gap.
%
%
Regarding our proposal, STM manipulation potentially has the advantage of simultaneously resolving the local density of states (LDOS).
Eventually, positioning the vortices amounts to engineering the hybridization between Majorana modes. Therefore, the presented finite-time results carry over to Y-junctions of Majorana wires \cite{Kitaev2001UnpairedMajoranaFermions,Alicea2011NonAbelianStatisticsIn1DWireNetworks,MourikKouwernhoven2012SignaturesOfMajoranaFermionsInHybridSCSMNanowireDevices,DasRonenMostOregHeiblumShtrikman2012ZeroBiasPeaksAndSplittingInAnAlInAsNanowireTopologicalSuperconductor}, where the hybridization between the Majorana modes at the periphery and the center is altered \cite{SauClarkeTewari2011ControllingNonAbelianStatisticsOfMajoranaFermionsInSCNanowires,MeidanGurRomito2019,KarzigOregRefaelFreedman2016UniversalGeometricPathToARobustMajoranaMagicGate,VijayFu2016}.

\paragraph{Setup}
We consider three vortex Majoranas $\gamma_1, \gamma_2, \gamma_3$ at the corners of an equilateral triangle with a fourth, movable vortex Majorana $\gamma_{\text{m}}$ near the center, see \refFig{figSetup}.
A key element of the setup is that the distances between the vortices minimize the hybridization between the exterior vortex Majoranas. 
%
%
%
%
In a low-energy, long-distance continuum model,
the hybridization strength of two vortex Majoranas 
 \cite{PhysRevLett.103.107001,PhysRevB.82.094504} is proportional to
$
\cos(k_F r + \pi/4) e^{- r/\xi}  /\sqrt{r}, 
$
where $r$ is the distance between their centers and $k_F$ the Fermi momentum.
Two vortex Majoranas hence decouple at distances
\begin{align}
\label{eqnDecouplingDistances}
r_j = \pi \left(j-3/4\right)/k_F, 
\end{align}
where $j$ is a positive integer.
The formula changes slightly for real systems and in the presence of multiple vortex Majoranas, yet negligibly as far as our results are concerned. Furthermore,  thermal fluctuations in the superconducting gap $\Delta$, which in principle set an upper limit to the temperature \cite{MishmashBauerOppenAlicea2019DephasingAndLeakageDynamicsOfNoidyMajoranaBasedQubitsTopologicalVersusAndreev}, can be neglected in the  regime of validity of \refEq{eqnDecouplingDistances}, which does not depend on $\Delta$.
%
%
In the setup (\refFig{figSetup}), the exterior vortices are $r_6$ away from each other, so that no hybridization takes place between them.
The central vortex is moved along a path that is exactly $r_4$ away from at least one exterior vortex, which results in the braiding path shown in \refFig{figSetup_path}(b). 
The setup has a $C_3$ symmetry and a mirror symmetry, which we assume to hold in the following if not stated otherwise.
Using $r_4$ and $r_6$ results in a particularly short braiding path on FeTe$_x$Se$_{1-x}$, where the superconducting Majorana coherence length is $\xi \approx 13.9$~nm. The direct distance between the exterior vortices is ${\sim}80$~nm, and $\gamma_{\text{m}}$ is never further than ${\sim}6.5$~nm away from the center. 
Other distances $r_j$ could perfectly be used as well, in particular for different material systems.

\paragraph{Braiding protocol}
In the following, we provide a step-by-step recipe for finite-time Majorana braiding using FeTe$_{0.55}$Se$_{0.45}$ as an exemplary platform.
To this end, we employ a material specific tight-binding model  \cite{2019arXiv190413374C,SupplementalMaterial}.
We discuss the essential steps 
\begin{inparaenum}
 \item arranging the vortex positions,
 \item calibrating the braiding path,
 \item performing the braiding gate, and 
 \item reading out quantum states. 
\end{inparaenum}

We first propose how to construct the setup shown in \refFig{figSetup}(a).  Consider vortex Majoranas on the surface of FeTe$_{0.55}$Se$_{0.45}$. In the low-field regime $\le 1$~T, the arrangement of the vortices is close to a triangular lattice \cite{Machida:2019aa}, where the strength of the magnetic field controls the lattice constant.
For the setup of \refFig{figSetup}(a), the tight-binding model predicts an ideal edge length of $r_6=77.3$~nm, corresponding to a magnetic field of $0.36$ T. The concrete experimental value can be obtained by calibration relative to this starting point. 
A deviation of about $0.1$~nm from the perfect positions does not crucially affect the protocol \cite{SupplementalMaterial}.
We next use the aforementioned techniques \cite{doi:10.1063/1.3000963,NovemberSauWilliams2019SchemeForMajoranaManipulationUsingMFS,PolshynNaibertBudakian2019ManupulatingMultiVortexStatesInSuperconductingStructures,Ge..Moshchalkov2016NanoscaleAssemblyOfSCVorticesWithScanningTunnelingMicroscopeTip} to move a vortex ($\gamma_{\text{m}}$) to the center of one Majorana triangle ($\gamma_1$, $\gamma_2$, $\gamma_3$)
and isolate this formation by moving surrounding vortices away. Finally, we precisely adjust the positions of these four vortices such that
only two zero-bias tunneling peaks appear at the positions of $\gamma_1$ and $\gamma_3$, while a small energy splitting of ${\sim}25$~$\mu$eV is present at $\gamma_2$ and $\gamma_{\text{m}}$, see \refFig{figSetup}. Because of this energy splitting, we assume a temperature below $0.3$~K for quantum information processing, which is achieved in state-of-the-art experiments.

The braiding path shown in \refFig{LDOS_evolution_new}(b) is calibrated by ensuring that the LDOS for each vortex Majorana during the braiding process qualitatively aligns with the results shown 
in \refFig{LDOS_evolution_new}(c).
Hence, at the beginning, only $\gamma_2$ and $\gamma_{\text{m}}$ hybridize so that $\gamma_1$ and $\gamma_3$ exhibit zero-bias peaks.
Furthermore, $\gamma_3$ always possesses a zero-bias peak during the first third of the braiding protocol.
When $\gamma_{\text{m}}$ moves towards $\gamma_1$, the energy splitting starts to transit from $\gamma_2$ to $\gamma_1$. The zero-bias peaks eventually appear at $\gamma_2$ and $\gamma_3$ when $\gamma_{\text{m}}$ is closest to $\gamma_1$.
In the remaining two thirds of the braiding path, the LDOS evolves as in the first third, but with suitably permuted indices of the vortex Majoranas, see \refFig{LDOS_evolution_new}(c).
Importantly, two vortex Majoranas remain at zero energy throughout the whole braiding process.
Furthermore, the energy splitting transits from $\gamma_2$ to $\gamma_1$ and then back to $\gamma_3$. On the other hand, the LDOS of $\gamma_{\text{m}}$ never possesses any zero-bias peak, as shown in \refFig{LDOS_evolution_new}(c), since it always couples to another vortex Majorana. Observing the LDOS evolution for each vortex is the primary step to check successful braiding.
To replace the technically demanding simultaneous measurement of the LDOS at each vortex, we first probe the LDOS at each vortex by an STM
and then move the central vortex $\gamma_{\text{m}}$ a short step along the braiding path. We iterate this procedure along the full braiding path until $\gamma_{\text{m}}$ is back to the starting point.

We next have to confirm the informational change
after braiding. 
The important data from the LDOS measurement to this end are the energy splitting and the peak heights of the vortex Majoranas that have been collected in the previous step, see \refFig{figSetup}(c). From the energy splittings, a low-energy model can be derived that determines the quantum gate operation on the degenerate ground state and its quality.
The model is explained in greater detail below. 
Braiding is then experimentally realized by moving $\gamma_{\text{m}}$ along the previously saved braiding path without probing any LDOS.

Finally, reading out the change of the quantum state after braiding is an important task, which has been discussed previously. In particular,
the Majorana qubit can be read out either by measuring the resonant current in Coulomb blocked systems\cite{LiuLiuZhangChiu2019ProtocolForReadingOutMajoranaVortexQubitAndTestingNonAbelianStatistics}, where charge fluctuations are suppressed and quantum information is thus protected, or by interferometry \cite{PluggeRasmussenAsbornEggerFlensberg2017MajoranaBoxQubits,ChiuVazifehFranz2015MajoranaShuttle,PekkerHouManucharyanDemler2013ProposalForCoherentCouplingOfMajoranaZeroModesAndSCQubitsUsingThe4PiJosephsonEffect}. 
Importantly, Majorana braiding and readout are distinct processes, such that the readout time is not limited to happen on the time scale of the braiding.

\begin{figure*}
\centering
\raisebox{-0.5 \height}{\subfloat[\label{figMajoranaCoupling}]{\includegraphics[height = 1 \figHeight]{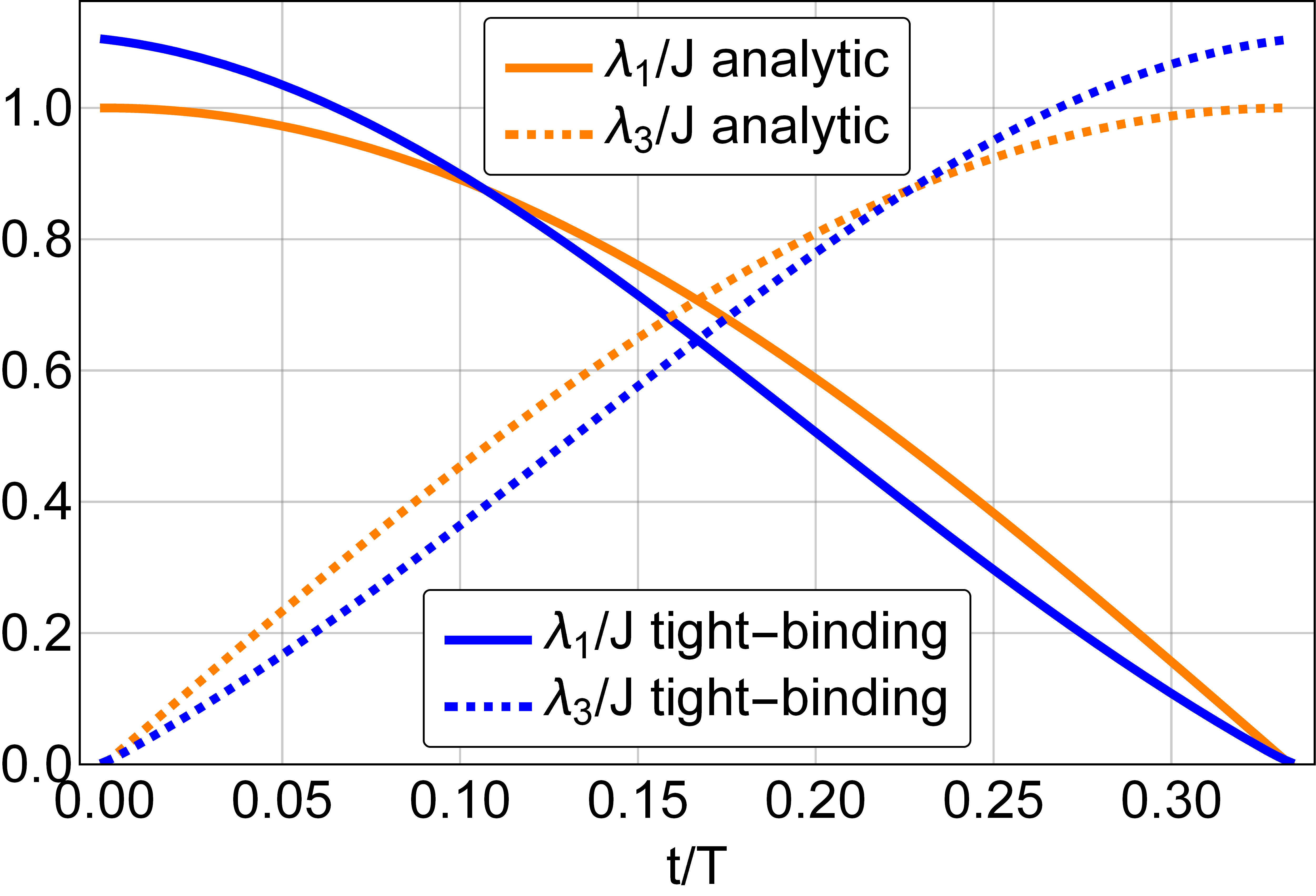}}}
\raisebox{-0.5 \height}{\subfloat[\label{figQuasiParticleExcitations3Steps}]{\includegraphics[height = 1 \figHeight]{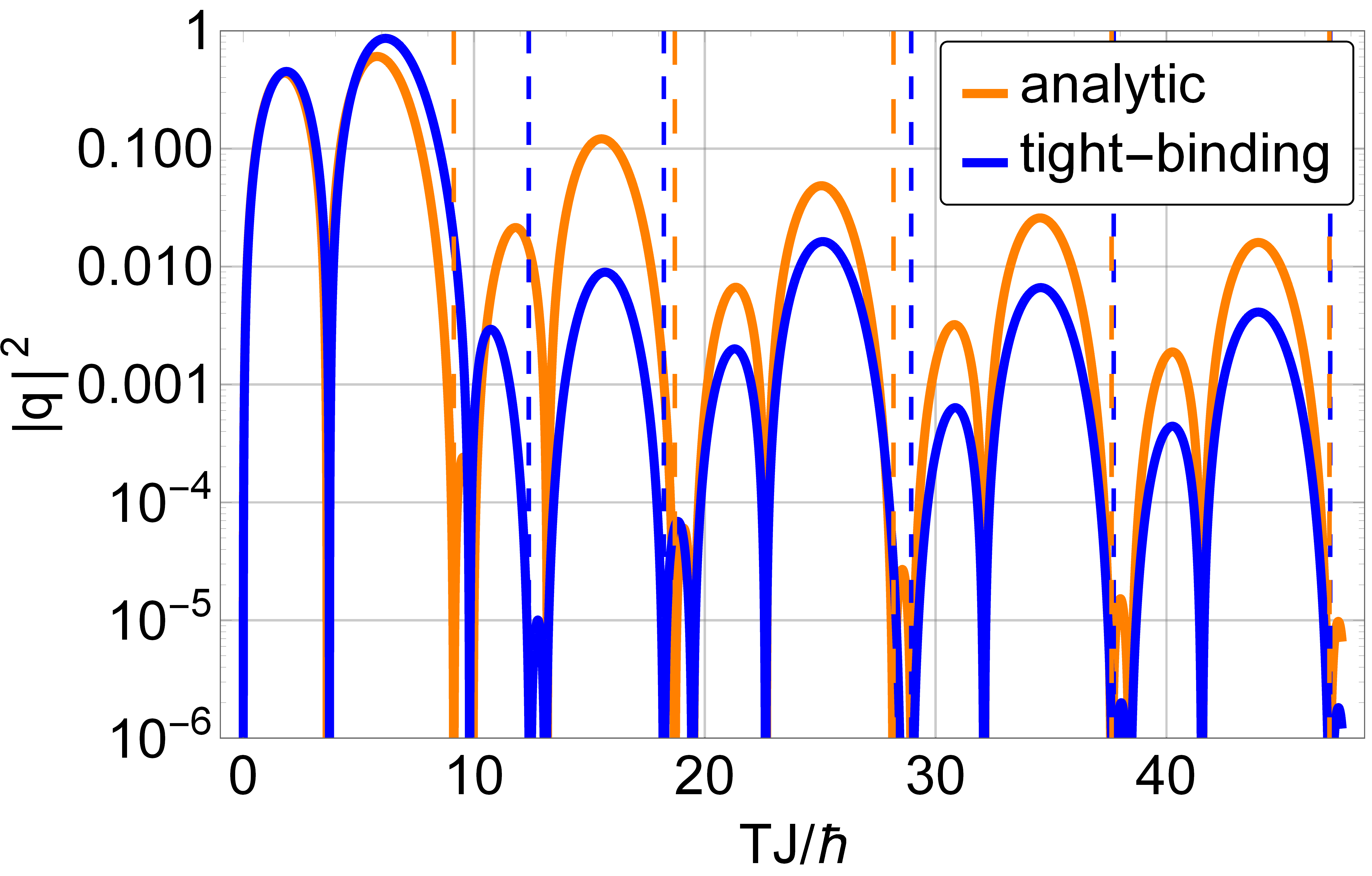}}}
\raisebox{-0.5 \height}{\subfloat[\label{figBerryPhase3Steps}]{\includegraphics[height = 1 \figHeight ]{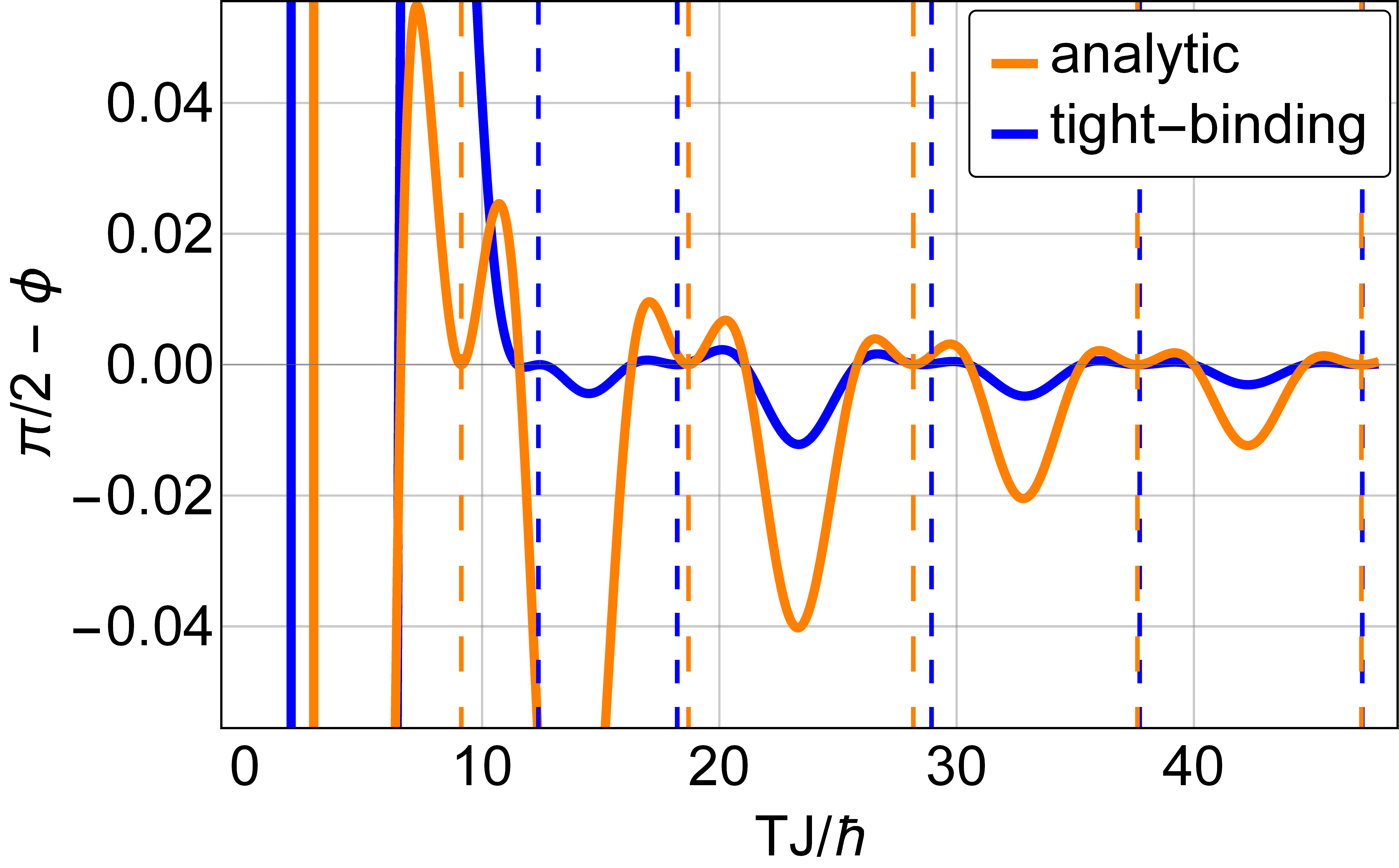}}}
\caption{\label{figComparisonAnalyticalAndTBModel}
Robustness of the braiding gate.
Realistic tight-binding parameters (blue) and analytically solvable parameters (orange) when $\gamma_{\text{m}}$ is moved along the braiding path in the time $T$.
{(a)} Hybridization strengths $\lambda_1$ and $\lambda_3$
($\lambda_2 = 0$ not shown) during the first third of the 
 braiding process. 
{(b)}
Unwanted excitations $|q|^2$ vanish around specific times $T_n$, \refEq{eqnSuperadiabaticTimes} (vertical dashed lines) for realistic as well as analytically solvable parameters.
The zeros are protected.
{(c)} When the quasiparticle excitations vanish at dashed lines, the finite-time Berry phase $\phi$ reaches the perfect adiabatic braiding value of $\pi/2$. %
}%
\end{figure*}


\paragraph{Braiding: validation, quality, and robustness}
To validate that Majorana braiding is realized, 
we consider the low-energy Hamiltonian
\begin{align}
\label{eqnHamiltonian}
\mathcal{H}(t) = \I J \left[\lambda_1(t) \gamma_1 + \lambda_2(t) \gamma_2+\lambda_3(t) \gamma_3 \right] \gamma_{\text{m}}\, ,
\end{align}
which describes the Majorana modes with energies much smaller than the one of the Caroli-de Gennes-Matricon (CdGM) states  
\cite{Wang..Gao2018EvidenceForMajoranaBoundStatesInAnFeBasedSC,Yin..Pan2015ObservationOfARobustZeroEnergyBoundStateInFeBasedSC}.
All $\gamma_j$'s obey the Majorana algebra \cite{Kitaev2001UnpairedMajoranaFermions}, $|\lambda_i(t)|\leq 1$ are time-dependent functions describing the hybridization strengths, and 
$J \approx 25$~$\mu$eV (for our setup on FeTe$_{0.55}$Se$_{0.45}$) is the maximal hybridization energy of two Majoranas.
Having used decoupling distances (\refEq{eqnDecouplingDistances} in-between most vortices, additional Majorana hybridizations are excluded.
Similar Hamiltonians have been studied in the adiabatic limit
\cite{Ivanov2001NonAbelianStatisticsOfHalfQuantumVorticesInPWaveSuperconductors,SauClarkeTewari2011ControllingNonAbelianStatisticsOfMajoranaFermionsInSCNanowires,MeidanGurRomito2019} 
and with projective measurements \cite{KarzigOregRefaelFreedman2016UniversalGeometricPathToARobustMajoranaMagicGate,VijayFu2016} in setups for superconducting wires. 
The distinguishing features of our work is that we employ a  time-dependent Hamiltonian and consider an experimentally realistic example system. 
Braiding on the time scale of GHz would tremendously outperform infinitely slow braiding. 

It can be shown directly that the braiding protocol works in the adiabatic regime. By the definition of the braiding path, and as shown in \refFig{LDOS_evolution_new}(c), one of the $\lambda_i(t)$ vanishes while two of the $\lambda_i(t)$ vanish at the corners of the triangle. 
The vector $\vec{\lambda}(t) = \left(\lambda_1(t), \lambda_2(t),\lambda_3(t)\right)$ therefore encloses a spherical angle of $\pi/2$ after the braiding protocol. In the adiabatic limit, the time evolution hence converges (up to a phase) to the braiding operator \cite{Kato1950OnTheAdiabaticTheoremOfQuantumMechanics,Berry1984QuantalPhaseFactorsAccompanyingAdiabaticChanges,MeidanGurRomito2019,SauClarkeTewari2011ControllingNonAbelianStatisticsOfMajoranaFermionsInSCNanowires,ChiuVazifehFranz2014MajoranaFermionExchangeInStrictly1DStructures} 
$
\lim_{T\to\infty} U(T) \propto 
1- \gamma_1 \gamma_2
\propto
B_{1,2}.
%
$

To describe the non-adiabatic regime, we fit the low-energy parameters of \refEq{eqnHamiltonian} to the realistic tight-binding model, see \refFig{figMajoranaCoupling}, allowing us to employ explicit Runge-Kutta methods for the numerical simulation of the time evolution. 
Because we assume $C_3$- and mirror symmetry to hold during the braiding process, we symmetrize the tight-binding data accordingly. Deviations from these symmetries can introduce small errors \cite{SupplementalMaterial}. 
To show that Majorana braiding is realized, we use two indicators.
These are the amount of quasiparticle excitations  $|q|^2$ and the phase difference $\phi$ between ground states of different parity.
In particular, $|q|^2$ is the probability of leaving the degenerate ground state by the finite-time manipulation of the vortex positions,
and the phase $\phi$ is an extension of the Berry phase to finite-time processes.
We neglect other sources of quasiparticle excitations, assuming a temperature below $0.3$~K, as stated above.
As shown in \refFig{figComparisonAnalyticalAndTBModel}(b) and (c), we find that the quasiparticle excitations $|q|^2$ and the deviation of the finite-time Berry phase $\phi$ from $\pi/2$ vanish simultaneously at times $T_n$.
Thereby, perfect finite-time braiding is achieved at times $T_n$ without physically braiding vortex Majoranas.
In FeTe$_x$Se$_{1-x}$ 
the corresponding time scales are larger than \mbox{$0.24$ ns}. 
This time scale is much slower than the time scale corresponding to excitations of CdGM states $\hbar E_F/\Delta^2 \approx 3.33$   ps, where $\Delta$ is the size of the superconducting gap and $E_F$ the Fermi energy.

The braiding protocol could be susceptible to variations in 
material parameters or the local speed of $\gamma_{\text{m}}$.  
In the following, we prove that these deviations are irrelevant by analytically solving a time-dependent Hamiltonian. 
We also find numerically that quasiparticle excitations stemming from slightly misplaced vortices are insignificant
\cite{SupplementalMaterial}.

If the time-dependent hybridization strengths along edge~$3$ of \refFig{figSetup_path}(b) take the form
\begin{align}
\label{eqnTrigonometricCouplings}
\lambda_1(t) = \sin\left({ \frac{3  \pi t}{2 T}}\right),\,
\lambda_2(t) = \cos\left({ \frac{3 \pi t}{2 T}}\right),\, 
\lambda_3(t) = 0,
\end{align}
the time evolution along edge~$3$ can analytically be given as
$U_{2,1}$, where
\begin{align}
\label{eqnExactTimeEvolution}
U_{j,k}(t) =&  e^{-\gamma_j \gamma_k \frac{3 \pi t}{4T}}e^{ \gamma_j \gamma_{\text{m}} \frac{J t}{\hbar} + \gamma_j \gamma_k \frac{3 \pi t}{4T}}\, .
\end{align}
That \refEq{eqnExactTimeEvolution} is the solution of the 
Schr{\"o}dinger equation is verified in \citeRef{SupplementalMaterial}. The solution is obtained by a rotating-wave ansatz and by solving a time-independent differential equation.
Notably,
the exact time-evolution of \refEq{eqnExactTimeEvolution} is equivalent to adiabatic braiding 
$U_{i,j}(T_n)=B_{i,j}$ at times
\begin{align}
\label{eqnSuperadiabaticTimes}
T_n = 3\pi \sqrt{n^2-1/16}\, \hbar/J,
\end{align}
where $n$ is a positive integer.
The time evolution for moving $\gamma_{\text{m}}$ along all three edges is
$
U(T) = U_{3,2}(T) U_{1,3}(T) U_{2,1}(T).
$
Hence $U(T_n) \propto B_{1,3}$.
The analytic model therefore realizes analytically proved perfect braiding by moving $\gamma_{\text{m}}$ along a short path in a finite time. 

We next consider the case where $\gamma_{\text{m}}$ is on edge~$3$ and the hybridizations $\lambda_1(t)$ and $\lambda_2(t)$ differ from the analytic solution but still keep $\lambda_3(t) = 0$.  This is the case for the realistic tight-binding model.  
The general time evolution that respects mirror and $C_3$ symmetry is $U^{\text{g}}_{2,1}$ with
\begin{align}
\label{eqnGeneralUnitaryOperator}
U^{\text{g}}_{j,k}(T) = b_1(T) +  b_{2}(T) \gamma_j \gamma_{\text{m}} 
+  b_3(T) \gamma_{\text{m}} \gamma_k + b_4(T) \gamma_k \gamma_j,
\end{align}
where $b_i(T)$ are real coefficients. 
Additional terms are excluded by the reduced number of Majorana hybridizations and by mirror symmetry.
The complete time evolution operator along all three edges is 
\begin{align}
U^{\text{g}}(T) = U^{\text{g}}(T)_{3,2} U^{\text{g}}(T)_{1,3} U^{\text{g}}(T)_{2,1}.
\end{align}
The probability to excite quasiparticles after a full passage of $\gamma_{\text{m}}$ along the braiding path is $|q^{\text{g}}|^2$ with
\begin{align}
\label{eqnQuasiParticleExcitationsGeneral}
q^{\text{g}}(T) =& \Tr
    \left\{ 
            \left(\gamma_1+\I \gamma_2\right)
            \left(\gamma_3+\I \gamma_{\text{m}}\right)
            U^{\text{g}}(T)        
    \right\}/4
    \nonumber \\
     =& \sqrt{2} e^{\frac{\I \pi}{4}} \left(b_4(T) - b_1(T) \right)\left(1-\left(\Sigma_i b_i(T)\right)^2\right).
\end{align}
The amount of quasiparticle excitations is hence given by a product of real polynomials. 
For example, the analytical model of \refEq{eqnTrigonometricCouplings} results in
\begin{align}
\label{eqnAnalyticFormulaForQuasiParticleExcitations}
q^{\text{a}} =& e^{\frac{\I \pi}{4}}\frac{\pi^3 \sin(\frac{\theta}{2})}{2 \theta^3}
\left[  \frac{{J}}{\hbar\omega} \frac{\theta}{2\pi} \sin(\theta)- \cos(\theta) -\frac{{J}^2}{\hbar^2\omega^2}\right],
\end{align}
with 
$\theta = \pi \sqrt{{{J}}^2+\hbar^2\omega^2}/\left(\hbar \omega\right)$ and $\omega = 3\pi / ( 4T ) $.
At times $T_n$, see \refEq{eqnSuperadiabaticTimes}, and additional transcendental times, the quasiparticle excitations $|q^{\text{a}}|^2$ hence vanish.
In real systems, $q^{\text{}}$ generally differs from the analytical solution, but remains a product of the real polynomials $b_1-b_4$, and $1 - (\sum_i b_i)^2$, each of which is continuously connected to its counterpart in the analytical solution.
Single zeros of $q^{\text{}}$ are therefore  shifted but not lifted by small deviations. 
Only large deviations eventually annihilate two zeros simultaneously.
In particular, the low-energy model extracted from the realistic tight-binding calculations deviates significantly from the analytically solvable protocol, cf. \refFig{figMajoranaCoupling}. Yet, the zeros of $q$ persist and only shift slightly as shown in \refFig{figQuasiParticleExcitations3Steps}.

To finally analytically verify perfect braiding, we consider the finite-time extension of the Berry phase given by the phase difference 
\begin{align}
\phi &= \arg\left\{
\langle 0 | U^{\text{g}}|0\rangle
/
\langle 1 | U^{\text{g}} |1\rangle
\right\}
\end{align}
between the states $|0\rangle$ and $|1\rangle = \left(\gamma_2-\I \gamma_{\text{m}} \right)|0\rangle$.
In the adiabatic limit, $\phi$ equals the Berry phase, which is $\pi/2$ for Majorana braiding.
We find that $\phi = \pi/2$, i.e., perfect braiding, at the zeros of $q$ where $b_1(T) = b_4(T)$ in \refEq{eqnAnalyticFormulaForQuasiParticleExcitations}, as shown in \refFig{figBerryPhase3Steps}. 
These are exactly the zeros corresponding to $T_n$ of \refEq{eqnSuperadiabaticTimes} in the analytical model.

\paragraph{Conclusions.}
We show that Majorana braiding with superconducting vortices can be achieved robustly and in finite time by only slightly moving the vortices. 
The procedure avoids a long-time, incoherent, physical braiding process. 
We simulate the protocol in a realistic, material-specific tight-binding model and prove its robustness against variations of  material parameters and a nonconstant braiding speed by an analytically solvable time-dependent model. 
Perfect braiding without physically braiding Majoranas therefore becomes possible in systems, where the superconducting coherence length at the surface $\xi$ is comparable to the Fermi wave length. This requirement is met by FeTe$_x$Se$_{1-x}$.
If $\xi$ is much larger, controlled vortex manipulation becomes impractical, whereas if $\xi$ is much smaller, the Majorana hybridizations fall below current experimental resolution. 
The finite-time braiding ultimately relies on tuning the coupling between Majorana modes. 
Therefore, the scheme can also be realized in Y-junctions of 1D topological superconductors or by inserting and moving magnetic or nonmagnetic adatoms in between vortex Majoranas. Alternatively, the positions of anomalous vortices carrying Majorana modes  \cite{FanEtAl2020ReversibleTransitionBetweenYSRStateAndMZMByMagneticAdatomManipulation} could directly be manipulated.

\begin{acknowledgments} 
We thank Bela Bauer, Tetsuo Hanaguri, Lingyuan Kong, Dong-Fei Wang, and Roland Wiesendanger for helpful discussions. T.P.\ and M.T.\  acknowledge support by the German Research Foundation Cluster of Excellence ``Advanced Imaging of Matter'' of the Universit{\"a}t Hamburg (DFG 390715994). C.-K.C. is supported by the Strategic Priority Research Program of the Chinese Academy of Sciences (Grant XDB28000000).
\end{acknowledgments}


\bibliographystyle{apsrev4-1}

%

\clearpage
\appendix
\onecolumngrid

\section{Appendix}

\renewcommand{\thefigure}{S\arabic{figure}}
\renewcommand{\thetable}{S\Roman{table}}
\renewcommand{\theequation}{S\arabic{equation}}
In this Supplemental Material, we describe the  realistic tight-binding model  (\refSctn{sctnRealisticTightBindingModel} 1), discuss errors introduced by breaking the $C_3$ symmetry in the setup or by misplaced vortices (\refSctn{sctnUnwantedCouplingsAndBreakingC3Symmetry} 2),
and verify the analytical solution of the time evolution stated in the main text  (\refSctn{sctnExactTimeEvolution} 3).

\section{Realistic tight-binding model for the braiding gate}
\label{sctnRealisticTightBindingModel}

The distances given in \refEq{eqnDecouplingDistances}
of the main text, which describe when vortex Majoranas decouple, are derived from an approximation for the hybridization of exactly two vortex Majoranas. 
However, the structure at hand comprises four vortex Majoranas, so that any hybridization between two vortex Majoranas will unavoidably be affected by the presence of the remaining two. In addition, the presence of magnetic flux from additional vortices will alter the hybridization.
%
Thus, we employ a more realistic tight-binding model for FeTe$_{0.55}$Se$_{0.45}$ \cite{2019arXiv190413374C} with four vortices for our theoretical calculation. This model captures the essence of the topological superconductivity in FeTe$_{0.55}$Se$_{0.45}$ by including a surface Dirac cone with real, material specific parameters --- its chemical potential $\mu=5$~meV, the Fermi velocity of the Dirac cone $\nu_F=25$~nm$\cdot$meV, and the Fermi momentum $k_F=\mu/\nu_F=0.2$~nm$^{-1}$. Furthermore, we introduce an s-wave superconducting gap $(\Delta=1.8$~meV) \cite{Zhang182} to include superconductivity on the surface and add vortices carrying 
one magnetic flux quantum, obeying the London equations with a London penetration depth of $\lambda=500$~nm \cite{PhysRevB.81.180503,PhysRevB.82.184506}. In this setup, a vortex Majorana with zero energy arises at a vortex core on the surface of FeTe$_{0.55}$Se$_{0.45}$ \cite{2019arXiv190413374C}. Its characteristic length scales are given by the Majorana coherence length $\xi=\nu_F/\Delta=13.9$~nm and the oscillation length of the Majorana hybridization $\pi/k_F=15.7$~nm.

\begin{figure}  
\includegraphics[width = 0.55 \linewidth]{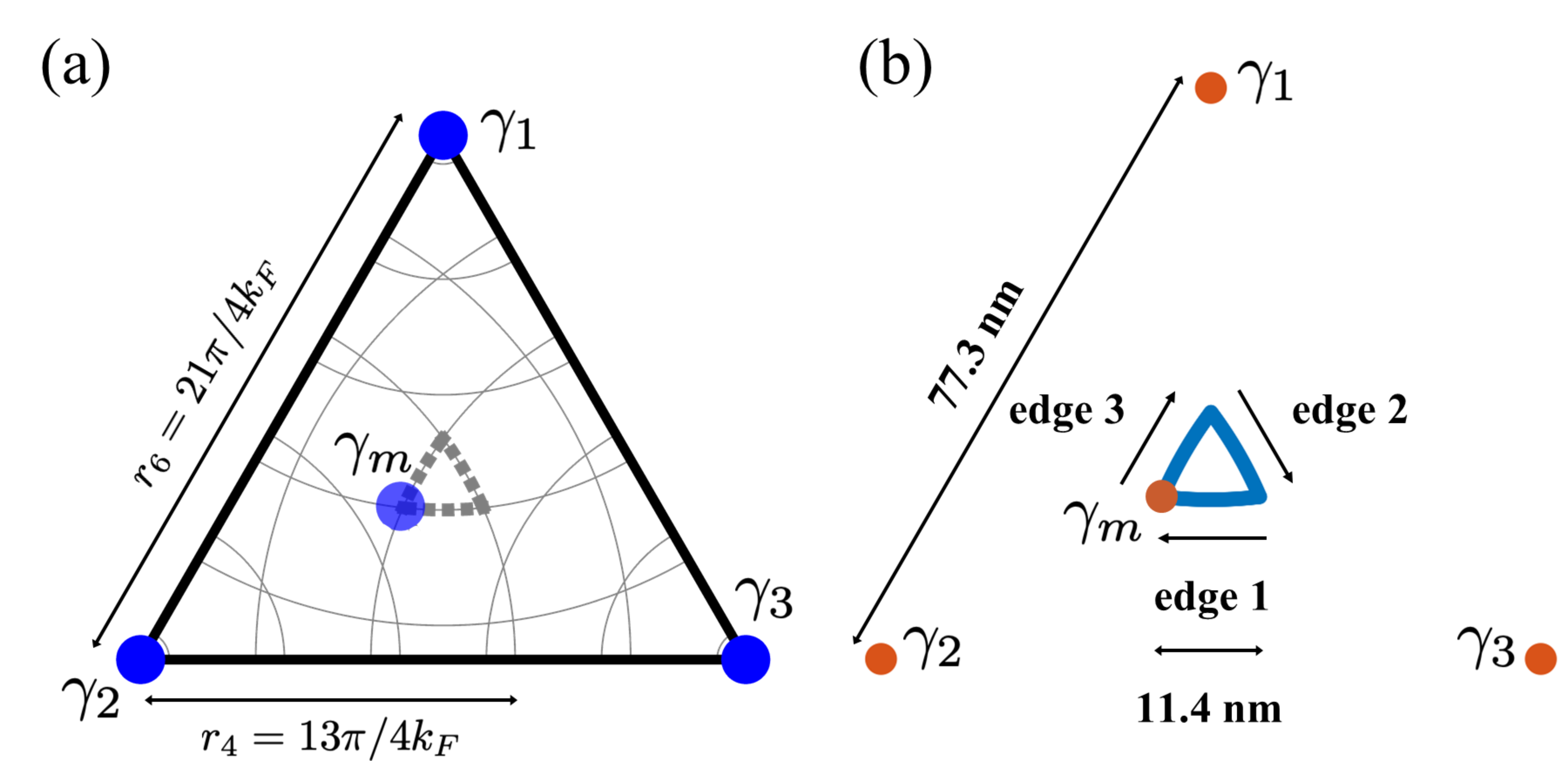}%
\caption{\label{Majorana_idea_braid}
The vortex configuration for setup in a long-distance continuum approximation {(a)} (cf. \refEq{eqnDecouplingDistances} of the main text)  and the resulting material specific length scales in the realistic tight-binding model {(b)}.
} 
\end{figure}

Next, we insert four vortices into our tight-binding model for FeTe$_{0.55}$Se$_{0.45}$ for the simulation of the braiding gate.
To find their correct positions, we start with the vortex distribution from the continuum model as shown in Fig.~\ref{Majorana_idea_braid}(a), with ($r_4=13\pi/(4k_F)=51.1$~nm, $r_6=21\pi/(4k_F)=82.5$~nm).
These distances are iteratively adjusted until the mediating vortex Majorana is able to move on the ideal braiding path.
%
At the start of the time evolution, the mediating Majorana is closest to the vortex Majorana $\gamma_2$.
%
To have the ideal initial conditions, we need $\I J\gamma_2\gamma_{\text{m}}$ to be the only hybridization surviving in the low-energy Hamiltonian 
\begin{align}
\label{eqnHamiltonianWithAllResidualCouplings}
H = \I \left(J \gamma_2 + a \gamma_1 + a \gamma_3 \right)\gamma_{\text{m}} + \I b \gamma_3 \gamma_1 + \I c \gamma_2 \left(\gamma_3 + \gamma_1\right),
\end{align}
whereas the other, residual hybridizations 
vanish.
Since additional factors (e.g.~magnetic flux, decay phase correction) influence the strength of the Majorana hybridization in the tight-binding model 
(as opposed to the analytical model),
the residual hybridizations generally do not exactly vanish. 
We need to find the optimal distances between the vortex Majoranas that minimize 
the residual hybridizations
by  solving the eigenvalue problem of the tight-binding model.
We find that, with a precision of $0.1$~nm,  an edge length of $77.3$~nm and a distance $\overline{\gamma_2 \gamma_{\text{m}}}=38$~nm is optimal. The minimal values of the residual hybridizations in the tight-binding model are $a/J=0.3\%$, $b/J=0.9\%$, and $c/J=1.4\%$, see \refEq{eqnHamiltonianWithAllResidualCouplings}. These residual Majorana hybridizations of about $10^{?2}J$ result from the derivation of the ideal braiding path, the positions of exterior vortices,  and other factors. 
After having found these optimal parameters, we start to move $\gamma_{\text{m}}$ towards $\gamma_1$.
That is, the hybridization between $\gamma_2$ and $\gamma_{\text{m}}$ gradually decreases and the hybridization between $\gamma_1$ and $\gamma_{\text{m}}$ increases as $\gamma_{\text{m}}$ is moved. To maintain the vanishing hybridization between $\gamma_1$ and $\gamma_{\text{m}}$, we keep the distance $\overline{\gamma_1\gamma_{\text{m}}}$ constant. 
We thereby assume that the hybridization strength $\I\gamma_1\gamma_{\text{m}}$ depends only on  the absolute distance, as it is the case for low-energy continuum model mentioned in the main text \cite{PhysRevLett.103.107001,PhysRevB.82.094504}.
When the mediating Majorana reaches the next corner of the braiding path, $\overline{\gamma_1\gamma_{\text{m}}}$ and $\overline{\gamma_2\gamma_{\text{m}}}$ have the same length and $\overline{\gamma_3\gamma_{\text{m}}}=38$ nm; the hybridization between $\gamma_1$ and $\gamma_{\text{m}}$ is dominant.
Following the same scheme, we move $\gamma_{\text{m}}$ towards $\gamma_3$ and finally return to $\gamma_2$. The conducted evolution is consistent with the one described in the main text.

\section{\texorpdfstring{Unwanted couplings and breaking of $C_3$ symmetry}{Unwanted couplings and breaking of C3 symmetry}}
\label{sctnUnwantedCouplingsAndBreakingC3Symmetry}
\begin{figure}
\includegraphics[width = 0.3 \linewidth]{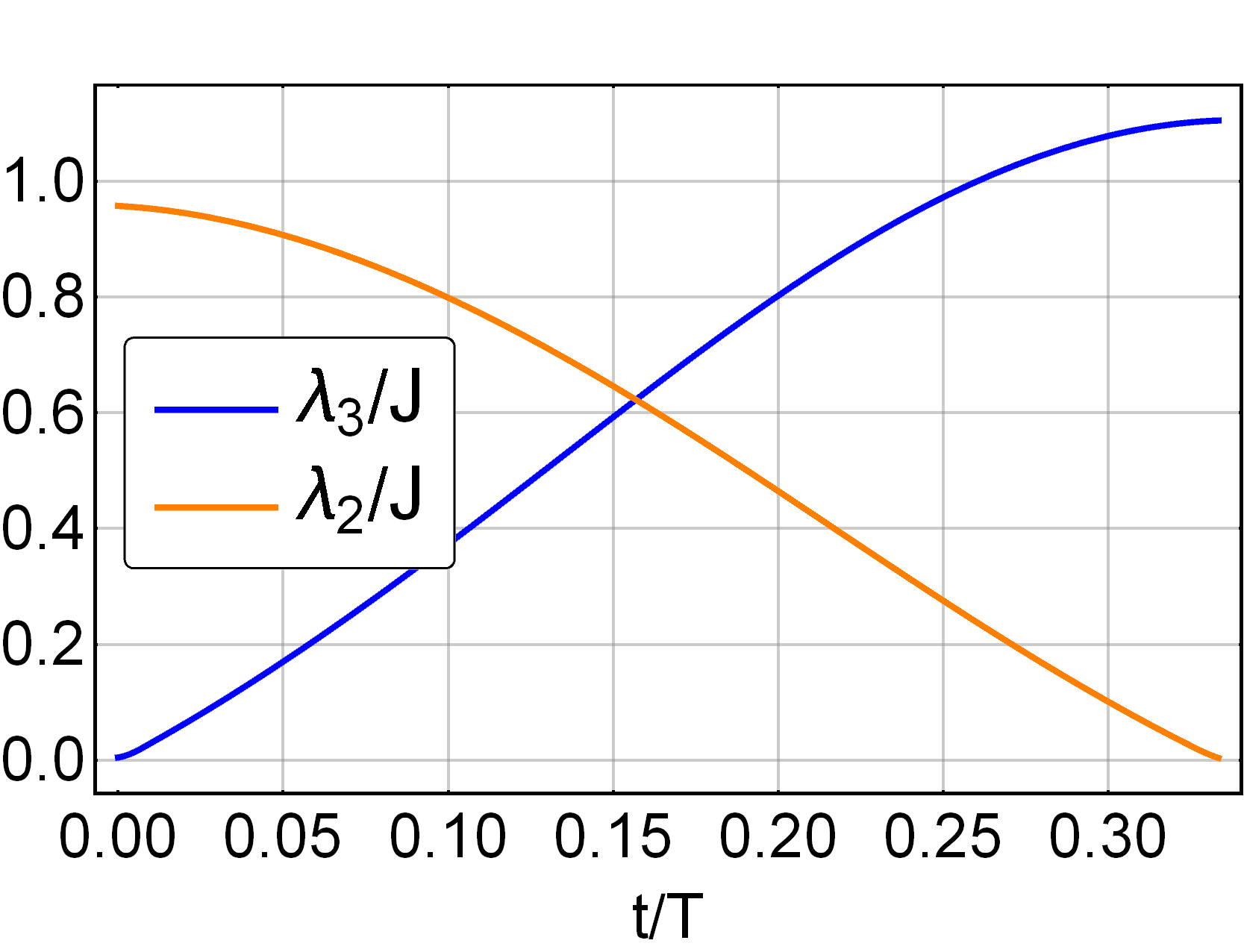}
\includegraphics[width = 0.3 \linewidth]{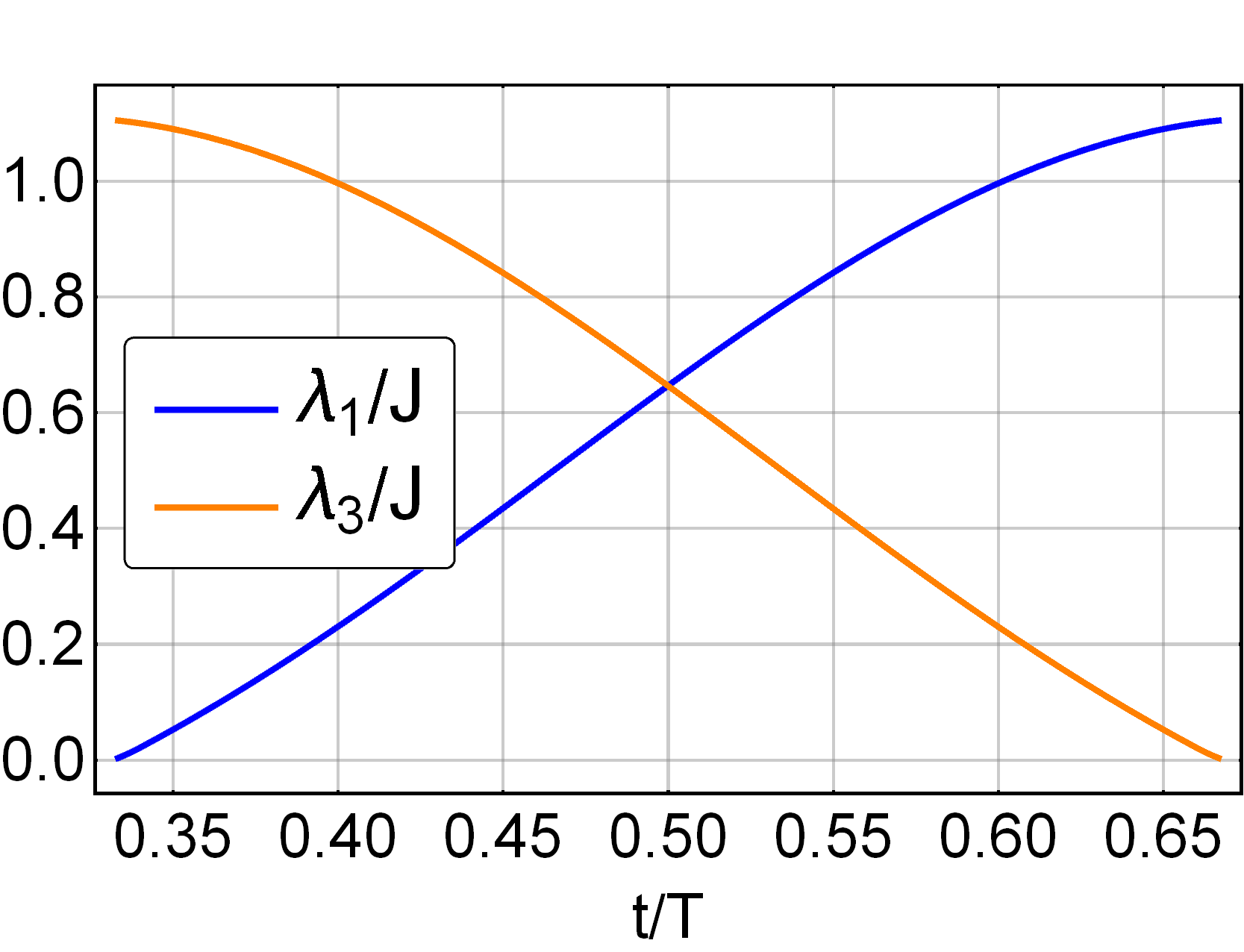}
\includegraphics[width = 0.3 \linewidth]{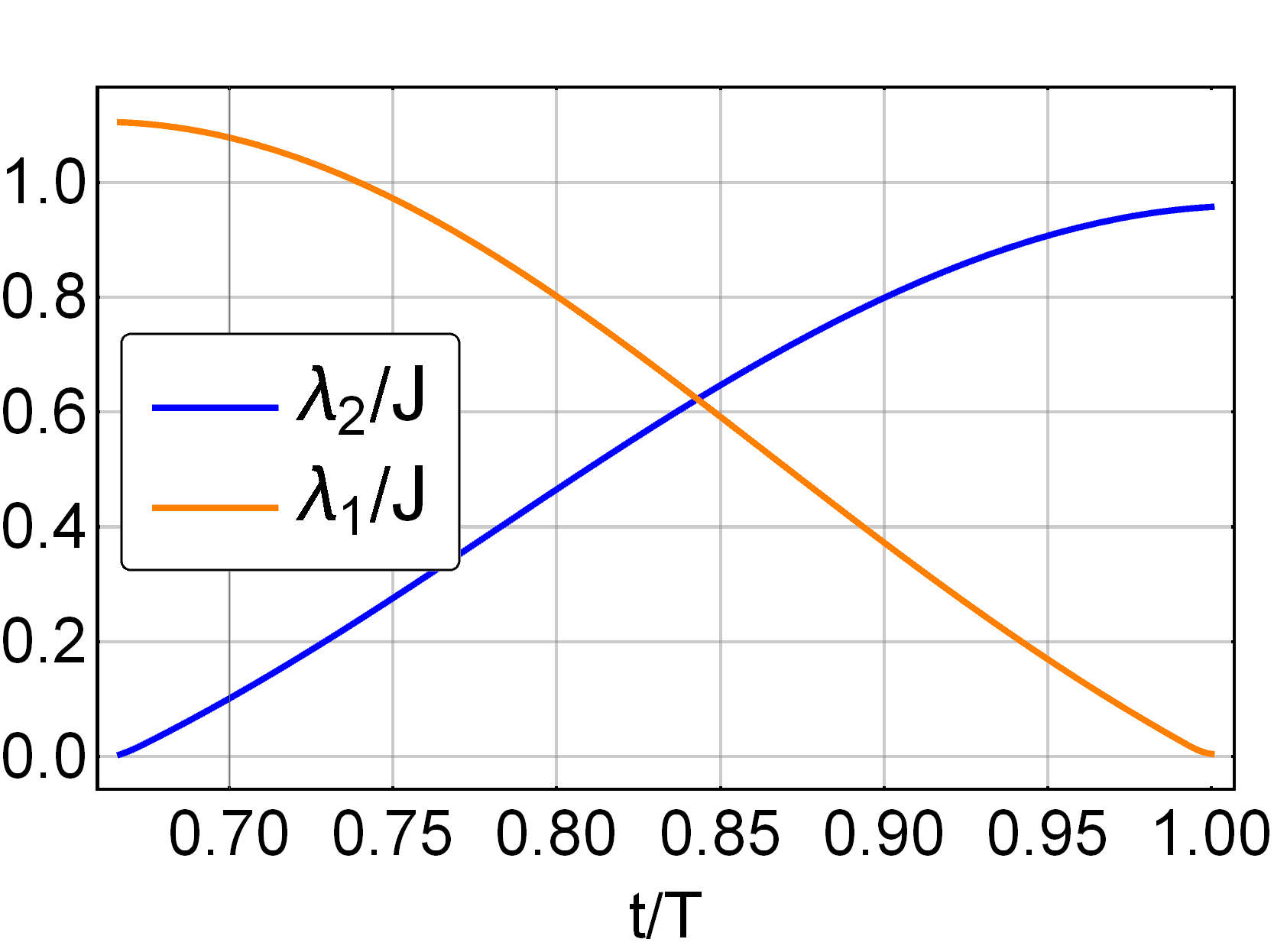}
\caption{\label{figC3BreakingCouplings}Realistic time-dependent  hybridizations that break the $C_3$ symmetry, extracted from the tight-binding model.
In each figure, the omitted coupling constant is close to zero.
}
\end{figure}

%
The proposed scheme is unexpectedly robust against variations in the residual couplings $a$, $b$, and $c$ ( see \refEq{eqnHamiltonianWithAllResidualCouplings}) as long as the $C_3$ symmetry of the setup, cf. \refFig{figSetup} of the main text, is maintained.
In this section, we give numerical evidence how unwanted quasiparticle excitations and deviations in the phase difference $\phi$ (see main text) emerge if the $C_3$ symmetry is slightly broken or if residual couplings are included that lift the degeneracy of the zero energy subspace.
To this end, we employ the protocol shown in \refFig{figC3BreakingCouplings}, which slightly breaks the $C_3$ symmetry, and incorporate the unwanted residual couplings $a/J=0.3\%$, $b/J=0.9\%$, and $c/J=1.4\%$, see \refEq{eqnHamiltonianWithAllResidualCouplings}, which respect a $0.1$~nm deviation from the perfect positions of the vortices.

\begin{figure}
\raisebox{-0.5 \height}{\subfloat[\label{figC3BreakingQuasiparticleExcitations}]{\includegraphics[height = 35 ex]{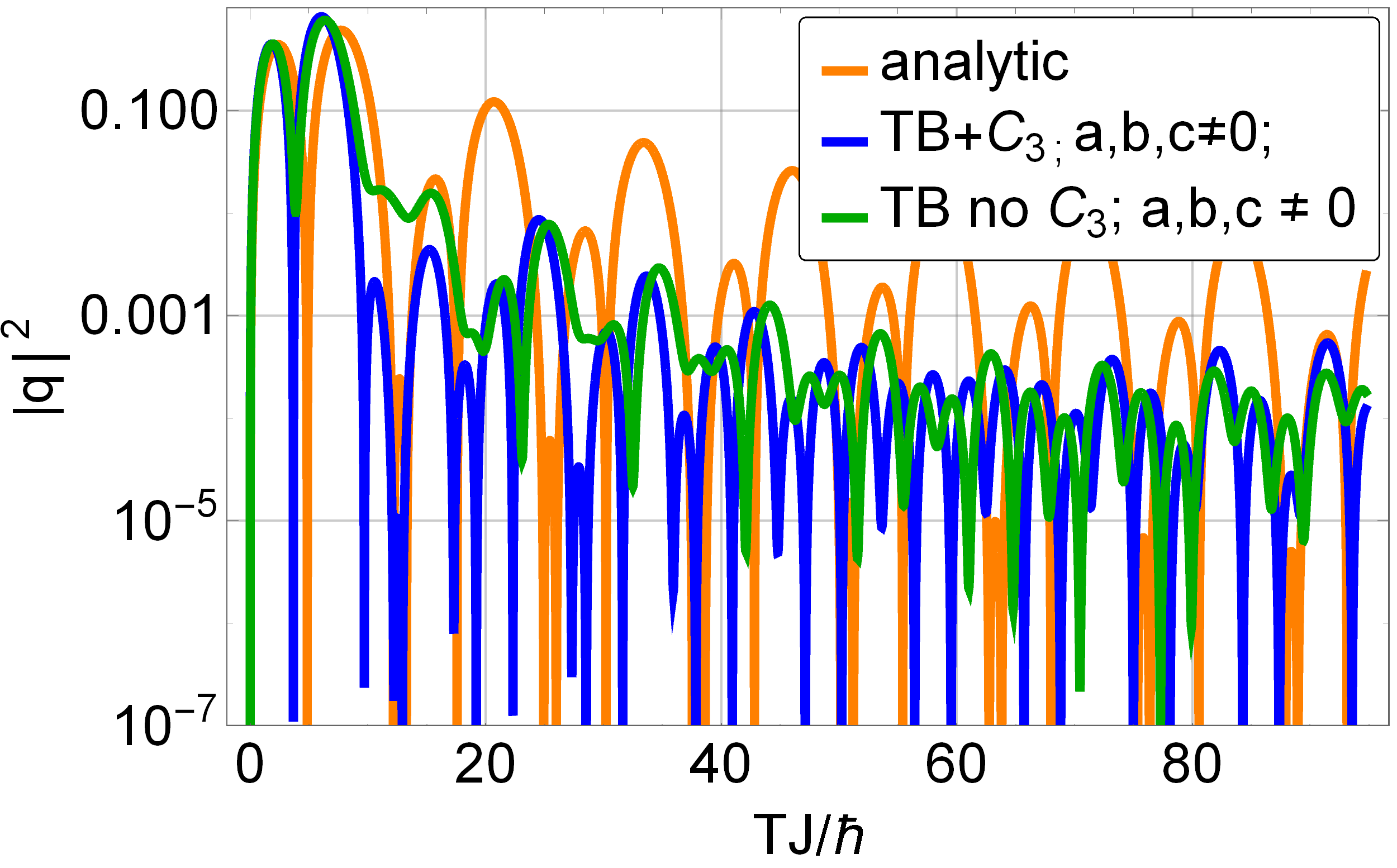}}}%
 %
\raisebox{-0.5 \height}{\subfloat[\label{figC3BreakingBerryPhase}]{\includegraphics[height = 35 ex]{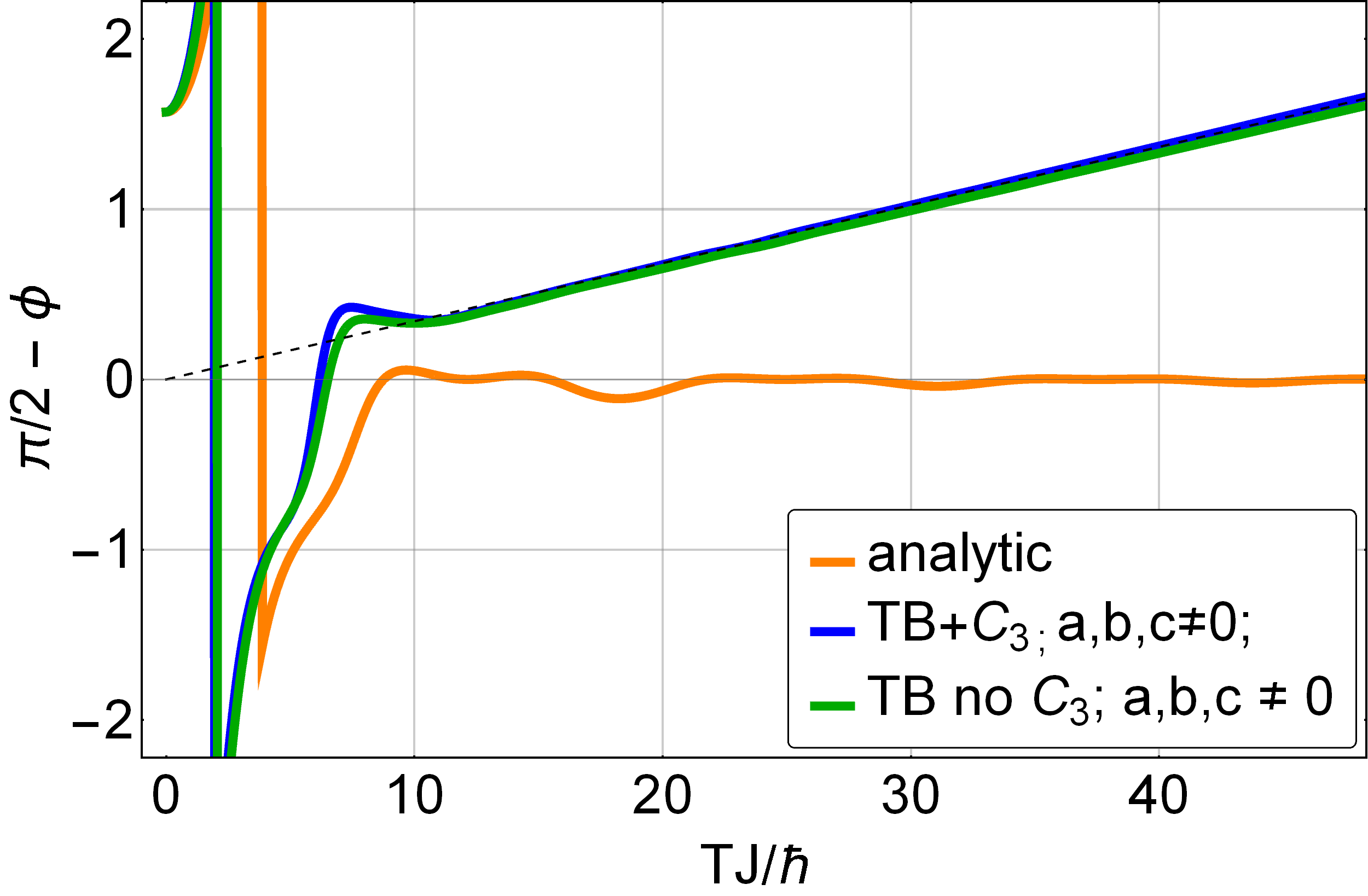}}}%
\caption{%
Realization of the braiding gate with additional realistic residual couplings (blue, numerical) and a broken $C_3$ symmetry (green, numerical).
The couplings are are $a/J=0.3\%$, $b/J=0.9\%$, and $c/J=1.4\%$, see \refSctn{sctnRealisticTightBindingModel} and \refEq{eqnHamiltonianWithAllResidualCouplings}.
For comparison, the analytic curves (orange) from the main text are included. {(a)} Unwanted quasiparticle excitations $|q|^2$ as defined in the main text. For a lifted ground state degeneracy (blue curve) there are times $T$ for which the unwanted quasiparticle excitations vanish. For an additionally broken $C_3$ symmetry (green curve), the formerly protected zeros lift.
{(b)} Finite-time Berry phase $\phi$ as defined in the main text. Since the ground state is no longer degenerate, an additional dynamic phase adds to the adiabatic Berry phase of $\pi/2$ (blue and green curve), which is approximately linear in time (asymptotic black, dashed line).
}
\end{figure}

As shown in \refFig{figC3BreakingQuasiparticleExcitations}, residual couplings are not increasing the unwanted quasiparticle excitations as long as the $C_3$ symmetry is not broken. Only breaking the $C_3$ symmetry results in additional quasiparticle poisoning, which can be suppressed to less than $10^{-6}$ by choosing braiding time.
Regarding the finite-time Berry phase, residual couplings introduce a time-dependent additional phase that needs to be compensated for, cf. \refFig{figC3BreakingBerryPhase}. This additional dynamical phase is expected, because unwanted couplings lift the ground state degeneracy  \cite{Kato1950OnTheAdiabaticTheoremOfQuantumMechanics,Berry1984QuantalPhaseFactorsAccompanyingAdiabaticChanges}. 
However, a deviation from the braiding phase of $\pi/2$ may  turn out to be a merit and not a deficiency. Each zero of the unwanted quasiparticle excitations corresponds to the realization of another phase gate. This has the advantage that different braiding speeds realize different phase gates (e.g., the $\pi/4$ magic phase gate).

%

%
%

\section{Verification of the analytic solution}
\label{sctnExactTimeEvolution}

In \refEq{eqnExactTimeEvolution} of the main text, we give the closed form of the time evolution corresponding to the Hamiltonian \refEq{eqnHamiltonian} with the time-dependent hybridizations of \refEq{eqnTrigonometricCouplings} of the main text.
Here, we verify the solution.
The time evolution operator is the solution of the Schr{\"o}dinger equation for the time evolution operator
\begin{align}
\label{eqnSchroedingerEquationAnalytic}
\partial_t {U}(t) = -\I / \hbar  H(t) U(t),
\end{align}
which is obtained from the usual form of the Schr{\"o}dinger equation $\partial_t \Psi(t) = -\I \hbar H(t) \Psi(t)$ by replacing $\Psi(t) = U(t) \Psi(0)$.
Here
\begin{align}
\label{eqnHamiltonianAppendix}
{H}(t) = 2 \I J \left[\cos\left(\frac{3\pi t}{2 T}\right) \gamma_3 +\sin\left(\frac{3\pi t}{2 T}\right) \gamma_1 \right] \gamma_{\text{m}}.
\end{align}
The solution to \refEq{eqnSchroedingerEquationAnalytic} can be found by a rotating-wave ansatz and the subsequent solution of a simpler differential equation. The result is
\begin{align}
\label{eqnExactTimeEvolutionAppendix}
U(t) =&  e^{-\gamma_3 \gamma_1 \frac{3 \pi t}{ 4 T}}e^{ \gamma_3 \gamma_{\text{m}} J t/\hbar + \gamma_3 \gamma_1 \frac{3 \pi t}{4 T}}
\nonumber \\
=&  e^{- \omega t \gamma_3 \gamma_1 }e^{\left(J/\hbar\right) t \gamma_3 \gamma_{\text{m}}  + \omega t \gamma_3 \gamma_1}
\nonumber \\
=& \left[\cos\left(\omega t\right) -  \gamma_3 \gamma_1 \sin\left(\omega t \right)\right]
\left[
\cos\left(\alpha t\right) +  t 
\sinc\left(\alpha t\right)\left( \left(J/\hbar\right) \gamma_3 \gamma_{\text{m}} + \omega \gamma_3\gamma_1 \right)
\right].
\end{align}
with $\omega = 3\pi/(4T)$ and $\alpha= \sqrt{\omega^2+(J/\hbar)^2}$.
To obtain the final form we have used the algebra of the Majorana operators. The calculation can be verified using the fact that the parity sectors decouple, and $e^{\I a \vec{n} \vec{\sigma}} = \cos(a) + \I \vec{n}\vec{\sigma}\sin(a)$, where $\vec{\sigma} = (\sigma_1,\sigma_2,\sigma_3)$ is a vector containing all Pauli matrices and $\vec{n}$ is a vector of length $1$.

Here, we verify the solution given above and in the main text by using a specific (faithful) representation of the operators, namely
\begin{align}
\gamma_1 =& \frac{1}{{2}}\left(
\begin{array}{cccc}
 0 & 0 & 1 & 0 \\
 0 & 0 & 0 & 1 \\
 1 & 0 & 0 & 0 \\
 0 & 1 & 0 & 0 \\
\end{array}
\right),
\gamma_2 =& \frac{1}{{2}\I} \left(
\begin{array}{cccc}
 0 & 0 & 1 & 0 \\
 0 & 0 & 0 & 1 \\
 -1 & 0 & 0 & 0 \\
 0 & -1 & 0 & 0 \\
\end{array}
\right) ,
\gamma_3 =& \frac{1}{{2}}\left(
\begin{array}{cccc}
 0 & 1 & 0 & 0 \\
 1 & 0 & 0 & 0 \\
 0 & 0 & 0 & -1 \\
 0 & 0 & -1 & 0 \\
\end{array}
\right) ,
\gamma_{\text{m}} =& \frac{1}{{2}\I} \left(
\begin{array}{cccc}
 0 & 1 & 0 & 0 \\
 1 & 0 & 0 & 0 \\
 0 & 0 & 0 & -1 \\
 0 & 0 & -1 & 0 \\
\end{array}
\right).
\end{align}
We insert \refEq{eqnExactTimeEvolutionAppendix} into \refEq{eqnHamiltonianAppendix} and find that the matrix representation of both sides of \refEq{eqnExactTimeEvolutionAppendix} are the same. Explicitely, it is
\begin{align*}
\left(-\I/\hbar
H(t) U(t)\right)_{j,k} = \I J/\hbar {A}_{j,k}(t) \text{, and }
\left(\partial_t U(t)\right)_{j,k} = \I J/\hbar A_{j,k}(t).
\end{align*}
Here,
\begin{align*}
A_{1,1}(t) &= \cos(\alpha t) \cos(\omega t) + t \sinc(\alpha t) \left[ \I J/\hbar \cos(\omega t) - \omega \sin(\omega t) \right]
\\
A_{2,2}(t) &= 
                \cos(\omega t)\left[
                                    \I J/\hbar t \sinc(\alpha t) - \cos(\omega t)     
                                \right]
                + \omega t \sinc(\alpha t) \sin(\omega t)
\\
A_{4,1}(t) &= \cos(\alpha t) \sin(\omega t) + t \sinc(\alpha t) \left[ \I J/\hbar \sin(\omega t) + \omega \cos(\omega t) \right]
\\
A_{1,4}(t) &=  \cos(\alpha t) \sin(\omega t) + t \sinc(\alpha t) \left[ -\I J/\hbar \sin(\omega t) + \omega \cos(\omega t) \right]
\\
A_{2,3}(t) &= -A_{4,1}(t),\ A_{3,2}(t) = -A_{1,4}(t), \
A_{3,3}(t) = A_{1,1}(t) ,\ A_{4,4}(t) = A_{2,2}(t).
\end{align*}
Matrix elements that are not given vanish. The solution is hence verified.

\end{document}